\documentclass[twocolumn,lengthcheck,amsmath,amssymb,prl,showpacs]{revtex4}

\usepackage{graphicx}
\usepackage{dcolumn}
\usepackage{bm}
\usepackage{amsmath,bbm,amssymb,pifont}

\newcommand{\etal}{\textit{et al.}}

\newcommand{\rmd}{{\rm d}}
\newcommand{\eps}{{\epsilon}}
\newcommand{\mean}[1]{\left\langle #1 \right\rangle}
\newcommand{\avg}[1]{\left\langle #1 \right\rangle}
\newcommand{\abs}[1]{\left|#1\right|}
\newcommand{\nl}{L}
\newcommand{\Lnse}{L}
\usepackage{subfigure}

\begin{document}

\title{Scaling law and stability for a noisy quantum system}

\author{Mark Sadgrove$^{1}$} 
\author{Sandro Wimberger$^{2}$}
\author{Scott Parkins$^3$}
\author{Rainer Leonhardt$^3$}
\affiliation{
$^{1}$ CREST, Japan Science and Technology Agency, Kawaguchi, Saitama 332-0012, Japan \\
$^{2}$Institut f\"ur Theoretische Physik, Universit\"at Heidelberg Philosophenweg 19, 69120 Heidelberg, Germany \\
$^{3}$Physics Department, The University of Auckland, Private Bag 92019, Auckland, New Zealand
}

\date{\today}

\begin{abstract}
We show that a scaling law exists for the near resonant dynamics of cold
kicked atoms in the presence of a randomly fluctuating pulse amplitude.
Analysis of a quasi-classical phase-space representation of the quantum system
with noise allows a new scaling law to be deduced.
The scaling law and associated stability are confirmed by comparison with quantum simulations and 
experimental data.

\end{abstract}


\maketitle

Coherent quantum phenomena may now be routinely observed in ultra-cold neutral atoms
manipulated by light fields detuned from atomic resonance. The unprecedented control
of atomic dynamics afforded by these atom-optical techniques has impacted a 
number of fields significantly in the last decade. In practical terms, the 
realization of cold-atom fountain atomic clocks and atom interferometers is very 
important for precision measurements and metrology in general~\cite{Berman}. Other promising
applications include the manipulation of atoms in optical lattices~\cite{OberthalerReview} with possible applications to quantum computing~\cite{QC_OptLattice}.

Aside from such practical applications, atom optics has also offered the means to create
ideal experimental implementations of model systems, in particular, the 
quantum kicked rotor known in this realization as the atom optics kicked rotor.
The system and its variants has been studied by a number of 
groups worldwide \cite{AOKRexp,Lille,qres,Wimberger2005a,Sadgrove2005a} due to the ease of observing such quintessential quantum
phenomena as dynamical localization~\cite{Fishman} and dynamical quantum resonance~\cite{Izr}.
Recent interest in the quantum resonance phenomenon comes not only from a fundamental perspective, but also
from the useful features of the resonance behavior. For example it has been shown that the 
resonance peaks exhibit sub-Fourier resonance scaling~\cite{Wimberger2005a,Sadgrove2005a}, opening the 
possibility of faster than Fourier signal detection using the resonance phenomenon~\cite{Lille}.
Additionally, our work has great relevance to similar proposals for precision measurements of 
the atomic recoil frequency~\cite{Prentiss}.

The cloud hanging over all planned implementations of quantum technologies, is that of
\emph{decoherence}~\cite{H2006} -- interaction with environmental degrees of freedom which leads to irreversible loss
of phase coherence in quantum systems. In atom-optics systems, decoherence typically arises due to 
spontaneous emission and timing and amplitude fluctuations in lasers. Typically, decoherence must be treated
statistically, and its effect is only made plain by simulating quantum master-equations. However, in the
case of the quantum kicked rotor, some progress has been made in treating the response to spontaneous
emission decoherence through a quasi-classical scaling theory~\cite{WGF1}. In this case the dynamics
of kicked atoms near a fundamental quantum resonance, dependent ostensibly on four parameters (kick number,
strength, period, and spontaneous emission rate) is reduced to a stationary function of two scaled
time variables, with a closed analytical form. The presence of this scaling belies the fact that moderate noise typically
destroys quantum correlations and it might be thought that the scaling function in the presence of spontaneous
emission is an isolated case where decoherence is analytically tractable. However, here we show that 
a scaling exists in the same system in the presence of amplitude fluctuations. Most remarkably, 
the fundamentally quantum decoherence process can be visualised with a classical phase space picture here.
The noise changes the topology of the phase space in a way that makes clear which parameter regimes will 
exhibit robustness to decoherence.

It is important to note that amplitude noise induced destruction of quantum correlations has been proven
for non-quantum resonance conditions\cite{R2000}. This naturally leads to the assumption that 
away from exact quantum resonance, amplitude fluctuations will rapidly induce quantum decoherence.
The contrary was proved by a recent experiment \cite{Sadgrove2004a}, but the cause of the stability near quantum resonance has remained opaque. We derive in the following a thorough 
theoretical understanding of this robustness based on a semiclassical scaling approach. Our theory compares very well
with measurements of near-resonant motion.

Experimentally, we realize a kicked atom system with noise by overlapping an optical standing wave
with a sample of cold atoms and pulsing the potential periodically. The height of the potential
can be controlled by adjusting the optical power transmitted through an acousto-optic modulator.
The system with amplitude noise may be represented by the Hamiltonian~\cite{Graham}
\begin{equation} 
H(t') =\frac{p^2}{2} + k\cos(z)\sum_{s=0}^{t-1} 
 (1+R_s)\delta (t'/\tau-s)\;, 
\label{eq:ham} 
\end{equation}
where $p$ is the atomic momentum in units of $2\hbar k_L$,  
$z$ is the atomic position scaled by $2k_L$, $t'$ is time, and $t$ is the total number of kicks. 
Amplitude noise enters in the factors $R_s$ which are random numbers distributed uniformly
on the interval $[-\nl/2,+\nl/2]$, where $\nl$ is a noise level between 0 and 2.
The scaled kicking period $\tau$ is defined by the equation $\tau=8\omega_r T$, 
where $\omega_r=\hbar k_L^2/2M$ is the recoil energy. The kicking strength is proportional to the optical standing wave 
intensity, and its measured value  was $k \approx 4.3$  or $k \approx 2.8$ for the two separate sets of experimental data considered here. The kicking strength varied by about 
10$\%$ across the atomic sample.

In our experiments, a sample of cold Cs atoms was prepared in a standard magneto
optical trap (MOT)~\cite{Sadgrove2004a}. The atom ensemble had an initial width in momentum of up to
$\sigma_p/(2\hbar k_L) \approx 8$. They were released from the trap and exposed to either 5 or 20
periodic pulses of width $480 \rm \, ns$ from an optical standing wave detuned by 0.5 GHz from atomic resonance.
For the 20 kicks experiments (with $k\approx 2.8$) the presence of spontaneous emission at a rate of $~2.5\%$
per kick lead to a slight lifting and broadening of the resonance peaks. We corrected for the broadening 
by subtracting an additional small, empirically determined constant from the off--resonant energies in this case.
 Atoms were than allowed to evolve freely for $12 \rm \,ms$ before
applying the MOT beams and imaging the resultant fluorescence on a CCD camera.
In this way, the momentum distribution of the atoms was calculated allowing comparison
with theoretical predictions.
\begin{figure}
\includegraphics[width=\linewidth]{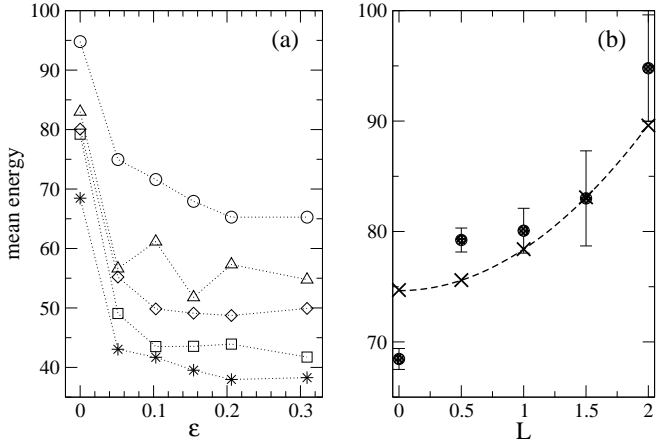}
\caption{\label{fig:peaks} (a) Experimental measurements of quantum resonance peaks
as a function of $\epsilon$ for noise levels $L=0$ ($\star$), $L=0.5$ ($\Box$), $L=1.0$ ($\diamond$),
$L=1.5$ ($\triangle$) and $L=2.0$ ($\circ$). (b) Points show experimentally measured peak energies (circles), 
whilst the dashed line shows the theoretical formula Eq.~\eqref{eq:Eresnse}. Crosses show
simulation energies, which exactly agree with the theory. Sample error bars are plotted from shot-to-shot measurements,
not taking into account systematic uncertainties in the absolute value of $k$.}
\end{figure}
It has been shown that for pulse periods $\tau$ equal to integer multiples of
$2\pi$ (so called fundamental quantum resonances) a semi-classical map may be
used to describe the quantum dynamics~\cite{WGF1}. We define a detuning $\epsilon = \tau - 2\pi\ell$ 
which measures how far the pulse period is from the $\ell$th fundamental quantum resonance,
and define new scaled momenta and position variables $J_s=|\epsilon|p_s + \pi\ell + \tau\beta$,
(where $p_s$ is the atomic momentum in units of $2\hbar k_L$ at kick $s$ and $\beta$ is the non-integer
quasi-momentum) and $\theta = z + \pi(1-{\rm sign}(\eps))/2 ~~{\rm \, mod}(2\pi)$. Then the pseudo-classical
standard map with amplitude fluctuations is (see \cite{WGF1,Sadgrove2004a})
\begin{equation}
\label{eq:ecsm_amp}
J_{s+1}=J_s+|\epsilon| k(1+R_s)\sin (\theta_{s+1})\;\;,\;\;
\theta_{s+1}=\theta_s+J_s\;.
\end{equation}
We now proceed to investigate how the mean energy at exact quantum resonance is affected by \emph{amplitude noise}.
To do this we need to find the average over all amplitude noise
realizations (and later initial conditions $\theta_0, J_0$) of the equation
$E_{t}(k_1,\dots,k_t) = \frac{1}{2\abs{\epsilon}^2}(J_{t}-J_0)^2
 \stackrel{\epsilon \to 0}{\longrightarrow} \frac{1}{2}\left[\sum_{s=0}^{t-1} k_{s+1}\sin(\theta_0+sJ_0)\right]^2$,
where we have used an expansion given in~\cite{WGF1}.
The noise average is given by  
$\mean{E_t(k_1,\dots,k_t)}_{R_j} =\prod_{j=1}^t\frac{1}{\nl}\int_{-\nl/2}^{\nl/2}\rmd R_j E_{t}(k_1,\dots,k_t).$
Since the series $\{k_s=k(1+R_s)\}$ is a series of independent random variables, this expression simplifies greatly.
Noting that $\mean{R_j}=0$ and $\mean{R_jR_i}=0,\;j\neq i$, we need only retain the following terms of $E_t$ in the integrand:
\begin{equation}
\label{eq:intexp}
\left[\sum_{s=0}^{t-1}\sin(\theta_0+sJ_0)\right]^2 + \left[\sum_{s=0}^{t-1}R_s\sin(\theta_0+sJ_0) \right]^2.
\end{equation}

We note in addition that 
$\langle[\sum_{s=0}^{t-1} R_s\sin(\theta_0+sJ_0)]^2\rangle
= \frac{\nl^2}{12}\sum_{s=0}^{t-1} \sin^2(\theta_0+sJ_0),$
where we have used
the fact that $(1/\nl^t)\int_{-\nl/2}^{\nl/2}\rmd R_1\dots\rmd R_s\dots\rmd R_t\;R_s^2
 = L^2/12$. 
Averaging over initial conditions $(\theta_0, J_0)$ gives, with $\theta \in [0,2\pi]$ and 
$J_0 \in [\pi \ell,\pi \ell + \tau]$ corresponding to a uniform quasi-momentum distribution 
in the unit interval (see \cite{WGF1}):
\begin{equation}
\label{eq:Eresnse}
\mean{\mean{E_{t,\nl}}} = 
 \frac{k^2}{4}t\left(1+\frac{\nl^2}{12}\right),
\end{equation}
where we have used the fact that the averages over both terms in Eq.~(\eqref{eq:intexp}) evaluate to $t/2$.
(This result was also given in ref.~\cite{B-Plata} from a purely quantum argument). Fig. 1 shows experimental data compared with simulation results and Eq.~\ref{eq:Eresnse}, demonstrating good agreement between all three.
 Shot to shot errors were found not to vary with $\epsilon$ and the 
given errorbars are estimates calculated from the standard error over 10 energy measurements at a kicking period of 
58$\mu s$. The discrepency between theory and experiment in the $L=0$ case is due to the difficulty
in measuring the high momentum components, a problem which is 
ameliorated by the addition of noise~\cite{qres,Sadgrove2004a}.

We now show how the scaling law introduced in \cite{WGF1} can be modified to take amplitude noise into account.
We start with the pseudo-classical scaling function~\cite{WGF1}
\begin{equation}
\label{eq:scale1}
\frac{\avg{E_{t,\nl,\epsilon}}}{\avg{E_{t,0}}} \equiv R(t,k, \epsilon) \approx H(x) \equiv 1 - \Phi_0(x) + \frac{4}{\pi x}G(x),
\end{equation}
where $x=t\sqrt{k\abs{\epsilon}}$ and $\avg{E_{t,0}} = k^2 t/4$ is the mean peak energy. The functions $\Phi_0$ and $G$ are 
evaluated numerically, and the reader is referred to ref.~\cite{WGF1} for details. 
\begin{figure}
\centering \includegraphics[width=\linewidth]{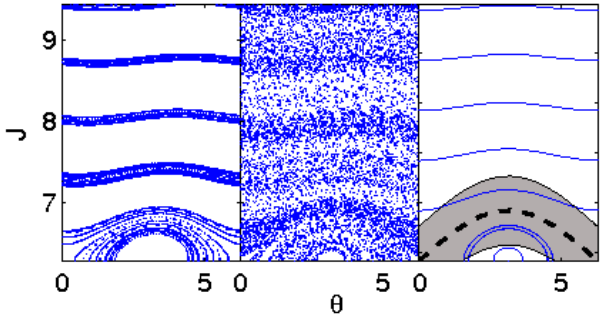}
\caption{\label{fig:phase} Phase space diagrams showing the effect of amplitude noise on the 
pseudoclassical map \eqref{eq:ecsm_amp}. Left panel: without noise. Middle panel: noise level $L=1.5$.
Right panel: pendulum trajectories for various initial conditions. The separatrix is shown by a thick, dashed line.
The grey shaded area shows a region of $\pm\mean{\Delta J_{\rm res}^2}$ for $L=1.5$ about the separatrix, demonstrating the trajectories which lead to the correction in \eqref{eq:scale_amp}.}
\end{figure}

For $\Lnse > 0$, we generally expect a loss of the scaling in all the variables $\epsilon, k, t, L$ due
to higher correlations in the evolution of the classical map \eqref{eq:ecsm_amp}, neglected above when deriving \eqref{eq:Eresnse}.
Remarkably, however, by observing the type of change in topology of the pseudo-classical phase space when increasing $\epsilon$,
as depicted in Fig.~\ref{fig:phase} we can nevertheless accurately estimate
the change of energy growth in the presence of noise for small $\epsilon$ (for which the semiclassical approach is
valid for long experimental time evolutions).
Noise is well known to enhance diffusion along nonlinear resonances in the first place \cite{LL92}. 
Therefore, we expect the major contribution of
energy enhancement around the separatrix region of pseudo-classical phase space, which separates the two different topologies
that give rise to the contributions $G$ and $\Phi_0$ to the scaling function~\cite{WGF1}. Since $G$ describes bounded librating pendulum motion
within the principal resonance zone, local changes of that motion due to noise will be small. The largest perturbation
comes from classical trajectories moving close to the separatrix which is washed out due to the fluctuations
of $k$ (see Fig.~\ref{fig:phase}). In this region, trajectories can actually perform rotating motion now, where at $L=0$ they would still be bounded to
the resonance. The increase of energy arising from those trajectories can be estimated by considering the area in phase space
covered by them, as shown for $L=1.5$ in Fig.~\ref{fig:phase}. Since the width of the principal resonance is given by $\Delta J_{\rm res} \approx 4 \sqrt{k\epsilon}$,
the relative {\em change} in weight of rotating orbits is given by
\begin{equation}
\label{eq:stima}
\frac{1}{2\pi}\frac{ \avg{\Delta J_{\rm res}^2} - \Delta J_{\rm res}^2 } { \Delta J_{\rm res}^2 } \approx \frac{L}{8\pi} \;.
\end{equation}
\begin{figure}
\centering \includegraphics[width=\linewidth]{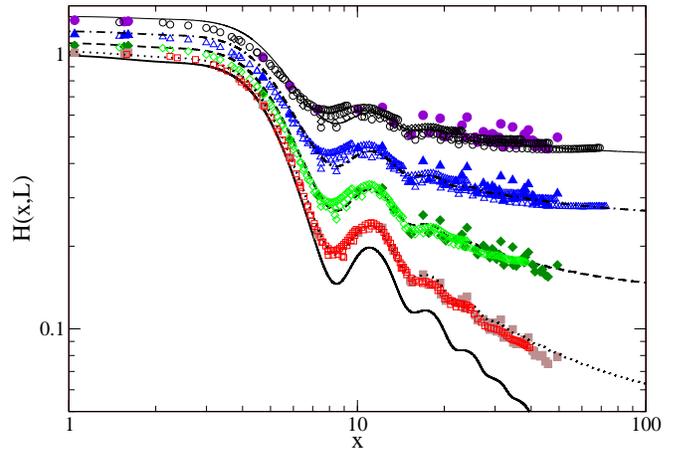}
\caption{\label{fig:scaledat} The theoretical scaling function for $L=0$ (thick solid line),
$L=0.5$ (dotted line), $L=1.0$ (dashed line), $L=1.5$ (dot-dashed) and $L=2.0$ (thin line).
Simulation data is also shown, rescaled by the factor $\frac{1}{4}k^2 t$,
for $\nl=0.5$ ($\Box$), $\nl=1.0$ ($\diamond$), $\nl=1.5$ ($\triangle$)
and $\nl=2.0$ ($\circ$). Open symbols are produced for fixed $k=2.8$, varying $\epsilon \in [10^{-3},0.1],
t \in [20,150]$, while filled symbols represent data for randomly chosen values of $k \in [1,10], 
\epsilon \in [10^{-3},0.1], t \in [2,150]$.}
\end{figure}
The noise-averaged standard deviation is $\avg{\Delta J_{\rm res}^2} \approx (1+L/4) \Delta J_{\rm res}^2$ by a simple integration.
With this result we can now add the additional energy of rotating trajectories to the scaling function from \eqref{eq:Eresnse},
by adding a term $L/(8\pi) \Phi(x)$. Dividing now the true energies by the result at exact quantum resonance 
and $L=0$, we finally arrive at the new scaling function for finite noise 
\begin{multline}
\label{eq:scale_amp}
\frac{ \avg{\avg{E_{t,\epsilon}}} } {\frac{1}{4}k^2 t} \equiv \mathcal{R}(t,\epsilon, k, L) \approx \mathcal{H}(x,\Lnse) \\
 \equiv 1 + \frac{L^2}{12} - \left( 1 - L/(8\pi) \right) \Phi_0(x) + \frac{4}{\pi x}G(x)\;.
\end{multline}
Our derivation of \eqref{eq:scale_amp} is thus analogous to the noise-free case, taking into account, however,
the main contribution of heating due to noise. 
Higher-order correlations and heating of the librating modes are neglected. We note that the principle changes to the
phase space which give rise to this scaling are readily seen in Fig~\ref{fig:phase}. In essentials, the scaling functiuon
reduces a complicated quantum system which includes decoherence to the dynamics of the pendulum.

\begin{figure}
\centering \includegraphics[width=\linewidth]{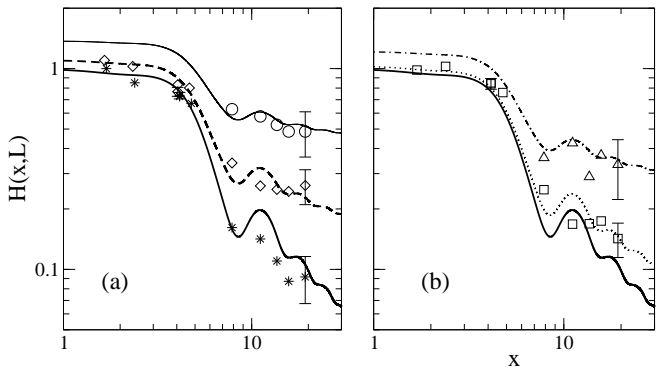}
\caption{\label{fig:scaleexpt} The theoretical scaling function of Eq.~\eqref{eq:scale_amp} (as shown in Fig.~\ref{fig:scaledat}) is compared with rescaled experimental data (as, e.g., from Fig.~\ref{fig:peaks}).
Shown are data across more than one order of magnitude in the scaling variable $x$ for (a)
$L=0$ ($\star$),  $L=1.0$ ($\diamond$), and $L=2.0$ ($\circ$) and (b) $L=0.5$ ($\Box$) and $L=1.5$ ($\triangle$). Theoretical
curves are shown with the same line-styles as in Fig.~\ref{fig:scaledat}.
Note that for $x < 5$ the data comes from separate 5 kick experiments, and the error bars are the same size as the plotted
points.
Sample error bars are calculated as described in the caption of Fig.~\ref{fig:peaks}. Data for $|\epsilon|>0.15$
is excluded since the pseudo-classical theory breaks down in this region.}
\end{figure}

Inspection of Eq.~\eqref{eq:scale_amp} reveals some interesting features as seen in Fig.\ref{fig:scaledat}. Firstly, because $\Phi_0$
saturates to 1 and $G(x)$ is small for small values of $x$, the small $x$ behavior is largely
unchanged in the scaling function. Essentially, the zero-noise scaling function is merely
displaced upwards for small $t$, $k$ or $\epsilon$. Experimentally, this means that as 
long as $x=t\sqrt{k|~\epsilon|}\lesssim4$ (e.g. take $t=20$, $k=0.1$ and scan over $\epsilon$ for any noise value),
 the resonance peak will not be broadened. This fact is important for proposed precision experiments such as~\cite{Prentiss}
where experimenters need to know how much tolerance the resonance width has to naturally occurring laser power
fluctuations.
Secondly, for large $x$ the scaling function is
significantly changed with the offset being much greater, corresponding to real broadening of the
peak and reduction of peak visibility. 

A comparison of the theory with simulation results is shown in Fig. \ref{fig:scaledat}. 
It may be seen that the scaling function reproduces the broad shape of the quantum simulations
over a large spectrum of parameters. Each point in Fig. \ref{fig:scaledat} is obtained by averaging
over 50,000 initial conditions, each of which is subject to kick-to-kick amplitude fluctuations.
Although our statistics are good, there is still a non-negligible scatter in the simulation data
which decreased systematically when augmenting the number of initial conditions averaged to obtain the final
energy. The experimental data from Fig.~\ref{fig:peaks} (a) and additional new data sets have been
plotted in Fig.~\ref{fig:scaleexpt}. The experimentally measured energies are obtained as
an ensemble average over the total number of atoms and are rescaled by subtracting the mean initial
energy of the ensemble $\sigma_p^2/4$ and then dividing by the energy at the peak maximum for $L=0$.
The estimated error bars shown in Fig.~\ref{fig:scaleexpt} represent shot-to-shot fluctuations over different noise realisations calculated as for Fig.~\ref{fig:peaks}.

In summary, we have derived and tested a generalized scaling function for the quantum resonance peaks in 
the presence of noise. The theory shows broad agreement with both quantum simulations and
experimental results. Most importantly it illuminates new facts about the response of quantum resonance to noise 
-- in particular, the stability of motion near to quantum resonance is revealed to be due to the unexpected
persistence of scaling laws in the noisy system. Although the effect of amplitude noise is to modify and even destroy quantum
correlations, the effect near to quantum resonance can be understood precisely in terms of the noise-induced
changes to the epsilon-classical phase space. Hence, quantum decoherence may be understood by a quasi-classical
analysis in the system studied here.
The robust nature of the scaling for small $x$ allows us to predict parameter families of $t$, $\tau$ and $k$ for which
noise will have a minimal effect on the quantum resonance, and surprisingly we find that for small enough  $x$,
the quantum resonance peak shape is entirely unaffected by noise (although a displacement in energy occurs).
The exploration of quantum systems which exhibit resistance to noise is of great importance for the future 
of quantum technologies. Analytical methods for determining the response of a system to noise and perturbations,
as done here and in a different context in \cite{Garreau}, are valuable because they offer insights on stability 
of quantum motion which simulations cannot readily provide.

The authors acknowledge support within the Excellence Initiative by the DFG through the Heidelberg Graduate 
School of Fundamental Physics (grant No. GSC 129/1) and thank Shmuel Fishman for stimulating discussions.

\end{document}